\documentclass[aps,prl,twocolumn,showpacs,amsfonts,amssymb,floats]{revtex4}
\usepackage{graphicx}
\usepackage{dcolumn}
\usepackage{bm}
\begin{document}
\noindent
{\bf Nonomura and Hu Reply:} 
In the preceding comment \cite{Olsson05}, Olsson and Teitel questioned 
the possible vortex slush (VS) phase in the frustrated XY model with 
point defects reported by the present authors \cite{Nonomura01}. The 
VS phase was originally proposed in order to explain an experiment of 
irradiated YBa$_{2}$Cu$_{3}$O$_{7-\delta}$ (YBCO) \cite{Worthington92}. 
This phase was also observed in an optimally-doped pristine 
YBCO \cite{Nishizaki00}, where the VS phase locates above 
the Bragg glass (BrG) phase in the $H$-$T$ phase diagram. 
In Monte Carlo simulations of the frustrated XY model 
by the present authors \cite{Nonomura01}, a first-order 
transition was observed between the vortex liquid (VL) 
and VS phases up to a certain density of point defects. 
The structure factor in the VS phase shows obscure Bragg 
peaks, which was interpreted as a short-range order 
in the $ab$ plane. In comparison with the BrG phase for 
lower density of point defects, the VS phase shows much 
larger density of dislocations in the $ab$ plane and 
the vanishing helicity modulus along the $z$ axis.

Olsson and Teitel simulated the same model for the 
same parameterization as used in Ref.\ \cite{Nonomura01}. 
On the basis of the structure factors observed in layer 
by layer, they argued that the VS phase observed by the 
present authors may be an artifact of a finite system size, 
and that this region may be included in the BrG phase in 
the thermodynamic limit. However, their argument in the 
Comment is not justified sufficiently by the provided 
numerical data for the following reasons.

First, they observed strong hysteresis behavior by sweeping the 
pinning strength of point defects $\epsilon$ in the VS region, 
and took this behavior as the evidence for a wide coexistence 
region of the first-order melting of the Bragg glass. However, 
this behavior can alternatively be interpreted as merging of 
the consequent VL-VS and VS-BrG first-order phase transitions 
due to a small system size. The two-step behavior in the 
hysteresis curve of the peak value of the structure factor 
in Fig.\ 1(b) of Ref.\ \cite{Olsson05} looks consistent with 
the latter picture. It should also be pointed out that the 
hysteresis behavior may be enhanced by their $\epsilon$-sweeping 
procedure. Since the annealing process is not included in 
this procedure in spite of possible drastic changes of the 
configurations of flux lines caused by varying $\epsilon$, 
Monte Carlo steps necessary for equilibration in this 
procedure may be much larger than those in the temperature 
sweeping adopted in our previous article \cite{Nonomura01}.

Second, they argued that the energy loss due to a mismatch 
in different layers is proportional to $J_{z}L^{2}$ with 
the transverse system size $L$. This scaling argument is 
based on the assumption that the mismatch characterized 
by the change of peak positions of Bragg peaks occurs 
as abruptly as a domain wall of the Ising model. However, 
the mismatches displayed in Fig.\ 2 of Ref.\ \cite{Olsson05} 
relax across a number of layers. When the relaxation takes 
place across $L_{w}$ layers, the energy loss is proportional 
to $J_{z}L^{2}/L_{w}$. It is natural to expect that $L_{w}$ 
depends on the thickness of the system $L_{z}$. Provided 
$L_{w}$ is proportional to $L_{z}\sim L$, the energy loss 
caused by a mismatch is proportional to $J_{z}L$, instead 
of $J_{z}L^{2}$. Then, their conclusion should be changed 
completely \cite{Remark}. In order to address the issue 
sufficiently, one should vary the system size and check 
the size dependence of the relaxation between mismatches 
and the number of mismatches.

On the other hand, we have to say that our previous 
study \cite{Nonomura01} cannot completely exclude the 
possibility that the ``first-order VL-VS phase boundary" may 
actually be the melting line of the finger-like wiggled BrG 
region stretching into the VL phase \cite{Li03}. Such a narrow 
BrG region (see Fig.\ 1 of Ref.\ \cite{Li03}) may be weakened 
by thermal fluctuations and the inverse-melting behavior may 
be masked by finite-size effects in numerical simulations or 
by some experimental conditions. Even if it is the case, the 
shape of the vortex phase diagram is qualitatively different 
from that obtained by Olsson and Teitel \cite{Olsson01}.

We also notice that the vortex phase diagram including the VS 
phase as Ref.\ \cite{Nishizaki00} was also reported in a YBCO 
thin film \cite{Wen01}, and that there exist several theoretical 
studies \cite{Kierfeld00,Ikeda01,Mikitik03,Rodriguez04} 
consistent with our numerical results.

In summary, the scaling argument which plays a key role in 
the preceding Comment cannot be sufficiently justified 
by the provided numerical data. Their data are indeed 
consistent with our previous article.
\bigskip
\par
\noindent
Yoshihiko Nonomura and Xiao Hu
\par
Computational Materials Science Center,
\par
National Institute for Materials Science,
\par
Tsukuba, Ibaraki 305-0047, Japan
\medskip
\par
\noindent
Received \today

\end{document}